\documentclass[sn-mathphys-num]{sn-jnl}

\usepackage[T1]{fontenc}%
\let\orcidlogo\relax
\usepackage{orcidlink}
\usepackage[utf8]{inputenc}%
\usepackage{graphicx}%
\usepackage{multirow}%
\usepackage{amsmath,amssymb,amsfonts}%
\usepackage{amsthm}%
\usepackage[title]{appendix}%
\usepackage{xcolor}%
\usepackage{textcomp}%
\usepackage{manyfoot}%
\usepackage{booktabs}%
\usepackage{soul}
\usepackage{tcolorbox}
\tcbuselibrary{listings,breakable}
\usepackage{geometry}
\begin{document}

\title[Structural Segmentation for Brazilian Legal Document Retrieval]{Structural Segmentation for Brazilian Legal Document Retrieval}

%%=============================================================%%
%% GivenName    -> \fnm{Joergen W.}
%% Particle -> \spfx{van der} -> surname prefix
%% FamilyName   -> \sur{Ploeg}
%% Suffix   -> \sfx{IV}

\author[1]{\fnm{Leandro} \sur{Carísio Fernandes} \orcidlink{0000-0002-4114-2334}}\email{carisio@gmail.com}

\author[2]{\fnm{João} \sur{Alberto de Oliveira Lima} \orcidlink{0000-0001-9830-5272}}\email{joaolima@senado.leg.br}

\author[2]{\fnm{Hudson} \sur{de Martim} \orcidlink{0000-0003-2751-3490}}\email{hudsonm@senado.leg.br}

\author[3]{\fnm{Eduardo} \sur{Seiti de Oliveira} \orcidlink{0000-0002-7882-6203}}

\author[3]{\fnm{Roberto} \sur{Lotufo} \orcidlink{0000-0002-5652-0852}}\email{lotufo@unicamp.br}

\author[3,4]{\fnm{Jayr} \sur{Pereira} \orcidlink{0000-0001-5478-438X}}\email{jayr.pereira@ufca.edu.br}

\affil[1]{\orgname{Câmara dos Deputados}, \orgaddress{\city{Brasília}, \country{Brazil}}}

\affil[2]{\orgname{Senado Federal}, \orgaddress{\city{Brasília}, \country{Brazil}}}

\affil[3]{\orgdiv{Faculdade de Engenharia Elétrica e da Computação}, \orgname{Universidade Estadual de Campinas (UNICAMP)}, \orgaddress{\city{Campinas}, \state{São Paulo}, \country{Brazil}}}

\affil[4]{\orgdiv{Centro de Ciências e Tecnologia}, \orgname{Universidade Federal do Cariri (UFCA)}, \orgaddress{\city{Juazeiro do Norte}, \state{Ceará}, \country{Brazil}}}

\abstract{
Long legal documents are often represented as single retrieval units, which can dilute relevant content and limit retrieval effectiveness. This study investigates structural segmentation as a document representation strategy to improve Brazilian legal document retrieval. Using the hierarchical structure of legal acts, we represent each document as a collection of chunks. We evaluate four representation strategies: a multi-layered hierarchy representation and three simplified versions based on core chunks (articles and non-normative document components), with varying contextual details. We test these strategies on two publicly available Brazilian legal datasets containing normative acts and legislative proposals, using both lexical and dense retrieval models. Results show that structural segmentation improves dense retrieval compared with whole-document indexing, despite substantially increasing the number of indexed units. However, it does not consistently benefit lexical retrieval. Finally, we find that full hierarchicy representation is not necessary: core chunks enriched with document metadata yield comparable or better results.}

%    \item Structural segmentation improves dense legal retrieval of Brazilian legal acts
%    \item Context-enriched chunks outperform context-free chunks
%    \item Core chunks achieve comparable results to full hierarchical representations

\keywords{legal information retrieval, document segmentation, dense retrieval, Brazilian legal acts}

\maketitle

\section{Introduction}\label{sec:introduction}

Recent advances in information retrieval (IR) have increasingly used dense vector representations. Unlike traditional lexical approaches, dense retrieval represents queries and documents as vectors, allowing semantic matching beyond exact word matches. However, long documents remain challenging to represent \cite{callan_passage, passage_retrieval_lm}. A single vector may provide a broad and imprecise representation of a document containing multiple topics, making relevant content harder to distinguish from unrelated information. One approach to this problem is to represent documents using smaller retrieval units, allowing the retriever to identify specific portions of a document that are relevant to a query.

This problem is relevant in legal information retrieval (legal IR), which applies IR techniques to legal documents. Legal documents differ from general-purpose text in several ways. They are often long, contain domain-specific terminology, and are highly structured, with relationships between their content and their organizational structure \cite{Moens2001}. Relevant information in a legal document may be concentrated in a single provision or section, while the remainder of the document may address different topics. Treating the whole document as a single retrieval unit can therefore result in an overly broad representation in which relevant information is diluted by unrelated content \cite{aumiller2021structural}.

The structure of legal documents provides an opportunity to address this problem. Structural components can be indexed separately, allowing retrieval models to match queries against more focused portions of the document. At the same time, smaller units may lose information needed to interpret their content. The choice of retrieval unit therefore involves a trade-off between representing localized information and preserving the context in which that information occurs. Legal documents are suitable for studying this trade-off because their structural organization is explicitly defined \cite{BHATTACHARYA2022103069}.

Brazilian legal documents provide a useful setting for this investigation. Legislative proposals, laws, and other normative acts follow formal drafting rules that organize their content into explicit structural units \cite{Brasil1998LC95, Brasil2003ResTCU164}. These units include articles and their subdivisions, as well as higher-level components such as sections and chapters. They provide natural boundaries for document segmentation, enabling retrieval representations based on the structure of legal acts.

The Brazilian legal retrieval context also presents challenges for current retrieval approaches. A recent benchmark of five Brazilian legal datasets showed that lexical retrieval models remained competitive with and, in some cases, outperformed dense retrieval models \cite{pereira2026juabenchmarkinformation}. The largest gaps occurred on NormasTCU and Ulysses-RFCorpus, two collections of highly structured legal documents \cite{normastcu, ulyssesrfcorpus}.

In this paper, we investigate structural segmentation of Brazilian legal acts for document retrieval. We represent each document as a set of structural chunks and evaluate different representation strategies. During retrieval, the model matches queries with individual chunks, and the resulting scores are mapped back to the original documents for document-level ranking. We evaluate these strategies on NormasTCU and Ulysses-RFCorpus using both lexical and dense retrieval models.

\subsection{Research Questions}

We address two research questions:

\textbf{RQ1}. Does structural segmentation improve Brazilian legal document retrieval compared with whole-document indexing?

\textbf{RQ2}. Which representation components contribute most to retrieval effectiveness?

\subsection{Contributions}

This work investigates the representation and retrieval of Brazilian legal acts. Its main contributions are as follows:

\begin{itemize}
    \item We investigate structural segmentation as a document representation strategy, comparing it with whole-document indexing using both lexical and dense retrieval models;

    \item We evaluate different structural representations by varying the structural chunks that are indexed and the contextual information associated with them;

    \item We show that context-enriched structural representations improve dense retrieval over whole-document indexing, while a simplified representation based on core chunks and document metadata can achieve comparable or better results than the full hierarchical representation.
\end{itemize}

\section{Related Work}\label{sec:related_work}

Passage-based retrieval addresses the limitations of whole-document representations by splitting documents into smaller textual units and scoring them independently. The passage scores can then be aggregated to produce a document-level ranking. A document can thus receive a high score when one or more of its passages are considered relevant to the query. Early studies showed that passage-level retrieval can outperform document-level retrieval \cite{callan_passage}. Later work has extended this approach through language models, probabilistic passage scoring, and alternative strategies for aggregating passage scores into document-level rankings \cite{parade, zhao2022densetextretrievalbased, Ai_2018}.

In Legal IR, passage-based retrieval has also been explored. Rosso et al. applied it to treaties, patents, and contracts in their experiments to find relevant sections rather than full documents \cite{rosso2011passage}. Aumiller et al. study the trade-off between document-level and smaller representations using a collection of English Terms of Service documents \cite{aumiller2021structural}. They argued that document-level representations may be too broad, whereas sentence- or paragraph-level representations may omit the contextual information needed to understand a topic. They proposed structural text segmentation based on topical coherence, identifying coherent sections that may span multiple paragraphs. Similarly, Malik et al. investigated semantic segmentation of legal documents using rhetorical roles \cite{malik2022semantic}. Their corpus consists of English-language documents from the Indian legal system, and their method identifies semantically coherent units corresponding to functions such as facts, arguments, issues, precedents, and rulings. Nguyen et al. investigated the retrieval of sets of legal articles related to a query or its negation in Japanese statute law, using individual legal articles as retrieval units rather than entire legal documents \cite{NGUYEN2025103949}.

Another approach uses structural information that is already present in legal documents. Structure-aware retrieval methods consider organizational or functional information to represent legal documents and their components. SAILER, for example, incorporates the structure of legal case documents into a pre-trained language model for legal case retrieval \cite{li2023sailer}. The authors argued that legal cases contain logical structures and key legal elements that are important for determining relevance, and they designed the model to exploit relationships between these structural components. Yang et al. also incorporated structured information into legal case retrieval, combining textual representations with structured factual representations based on event chains and participant roles \cite{YANG2026104816}. These studies show that document structure can provide information about how different parts of a legal document should be represented.

The importance of document structure is especially clear in legal acts that follow formal drafting conventions. In such documents, structure is part of how information is organized. Brazilian legal acts provide an example. Complementary Law No. 95/1998 establishes drafting rules for Brazilian laws, including rules concerning their organization and hierarchical structure \cite{Brasil1998LC95}. Other normative documents, including regulations, resolutions, ordinances, and legislative proposals, are organized according to similar principles (e.g. \cite{Brasil2024Dec12002, Brasil2003ResTCU164}). This formal organization provides explicit boundaries that may be used for document segmentation and retrieval.

Legal IR has developed around a range of specialized tasks and benchmarks, with much of the established literature focusing on judicial case retrieval and legal datasets from a limited number of jurisdictions. More recently, interest in Legal IR for Brazilian Portuguese has grown, with the development of publicly available datasets, retrieval pipelines, and benchmarking initiatives \cite{pipeline_cd, ulyssesrfcorpus, fernandes2026juristcu, brtax, pereira2026juabenchmarkinformation, normastcu}. These resources have enabled systematic comparisons of retrieval methods across different collections. However, relatively little work has considered the internal organization of Brazilian legal documents for retrieval \cite{lima2024unlockinglegalknowledgemultilayered, demartim2026probabilisticsimilaritystructuraltemporal}.

In the Brazilian context, Lima proposed a multi-layered representation for Brazilian legal acts \cite{lima2024unlockinglegalknowledgemultilayered}. Originally developed for Retrieval-Augmented Generation (RAG) applications, it exploits their hierarchical organization to represent each document at different structural levels. In this work, we investigate its use for Legal IR, where the goal is to retrieve the source document rather than individual chunks. We also conduct ablation studies to evaluate simplified representations derived from Lima's approach by varying which chunks are indexed and which contextual information is included with them.

\section{Methodology}\label{sec:methodology}

\begin{figure}[htbp]
    \centering
    \includegraphics[width=1\linewidth]{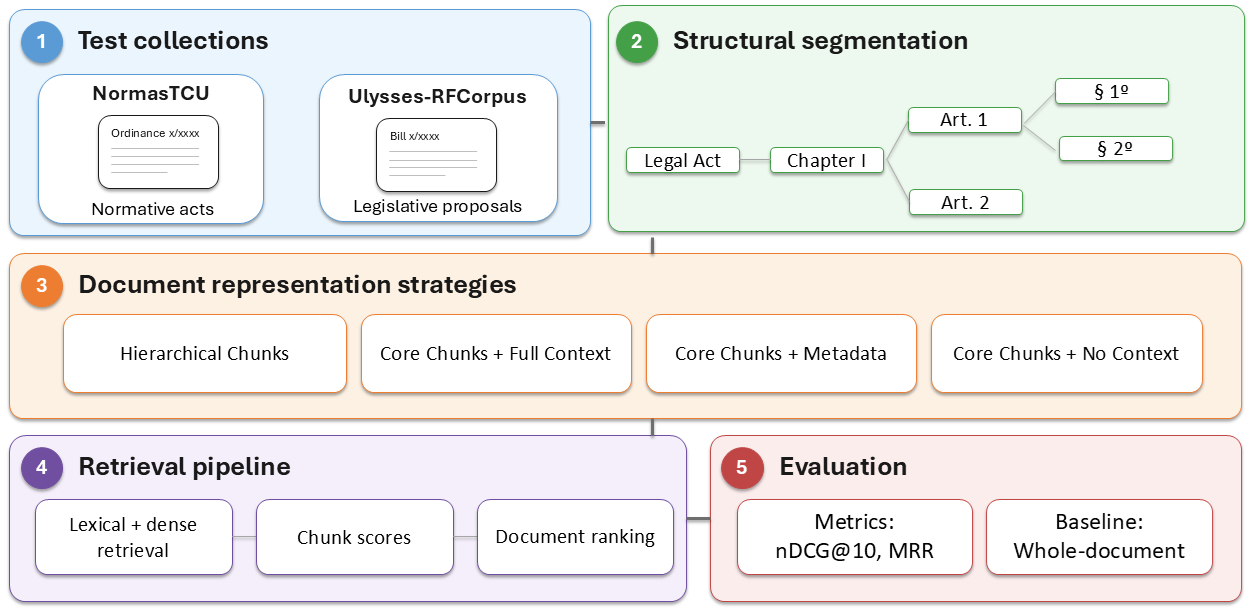}
    \caption{Overview of the proposed methodology.}
    \label{fig:methodology}
\end{figure}

The methodology is organized into five subsections. First, we describe the datasets used in the experiments. Second, we segment the documents according to their structural organization, representing each document as a collection of chunks. Third, we consider four document representation strategies based on these chunk collections, varying which chunks are indexed and what contextual information is included with them. Fourth, we index the chunks according to each representation strategy and retrieve them using lexical and dense retrieval models, mapping chunk scores to document scores. Finally, we evaluate the resulting document rankings using nDCG@10 and MRR against whole-document baselines. Figure \ref{fig:methodology} shows an overview of the methodology.

\subsection{Test collections}
\label{sec:collections}

We conduct the experiments on two public Brazilian legal test collections: NormasTCU and Ulysses-RFCorpus \cite{normastcu, ulyssesrfcorpus}.

NormasTCU contains 14,469 normative acts from the Brazilian Federal Court of Accounts. Its 46 queries are based on real user searches and mostly consist of short keywords. Query--document pairs were assessed by a panel of four domain experts using three relevance levels: irrelevant, partially relevant, or relevant

Ulysses-RFCorpus contains 105,669 legislative proposals from the Brazilian Chamber of Deputies. Each proposal also contains a justification section in which the deputy explains the motivation for the proposal, which is not present in the documents in NormasTCU. Its 693 queries correspond to requests made by deputies for legislative proposals. The relevance judgments were provided by 54 domain experts, with each query and its associated documents assessed by a single expert using three relevance levels: irrelevant, partially relevant, or relevant.

Both collections consist of raw text that generally follows the official legal drafting standards discussed in Section~\ref{sec:related_work} \cite{Brasil1998LC95, Brasil2003ResTCU164}.

\subsection{Strucutural segmentation}
\label{sec:doc_structural_segmentation}

Brazilian legal acts follow drafting rules that organize their content into explicit structural units. In these documents, articles (\textit{artigos}) are the main units of the normative text. Each article consists of a lead paragraph (\textit{caput}) and optional paragraphs (\textit{parágrafos}) that complement it. Each lead paragraph or paragraph may contain one or more levels of enumeration (\textit{inciso}, \textit{alínea}, and \textit{item}) \cite{lima2024unlockinglegalknowledgemultilayered}.

Articles can be organized into higher-level structural units, such as subsections (\textit{subseção}), sections (\textit{seção}), chapters (\textit{capítulo}), titles (\textit{título}), books (\textit{livro}), and parts (\textit{parte}). Together, these elements organize the main body of the normative document  \cite{lima2024unlockinglegalknowledgemultilayered}. Documents also have a title identifying their type, number, date of enactment, and a summary (\textit{ementa}). They may also contain other structural elements, such as tables, annexes, or a justification of the proposal (\textit{justificação}). 

In this work, we adopt the segmentation proposed by Lima \cite{lima2024unlockinglegalknowledgemultilayered}, which represents each document as a multi-layered collection of chunks at the following levels:

\begin{itemize}
    \item Document level: the whole text;
    \item Document component level: the main normative text and non-normative components, such as tables, annexes, and the justification of legislative proposals;
    \item Basic unit hierarchy level: each higher-level grouping of articles (e.g., subsection, section, chapter, title, book, or part);
    \item Basic unit level: each article, including all of its subdivisions;
    \item Basic unit component level: each lead paragraph (\textit{caput}) and each paragraph within an article; and
    \item Enumeration level: each individual enumerated item within a lead paragraph or paragraph.
\end{itemize}

Each chunk can be enriched with contextual information, including the document title, summary, and titles of all enclosing structural units (e.g., chapter, section, or subsection).

To evaluate structural segmentation, we focus on segmentation levels from the document component level to the enumeration level. Document-level retrieval has already been evaluated in a public benchmark and is used as a baseline for comparison \cite{pereira2026juabenchmarkinformation}.

To illustrate the segmentation method, Figure~\ref{fig:sample_bill} presents a sample legislative proposal, and Figure~\ref{fig:segmentation_sample_bill} shows its resulting structural chunks across hierarchy levels:

\begin{figure}[htbp]
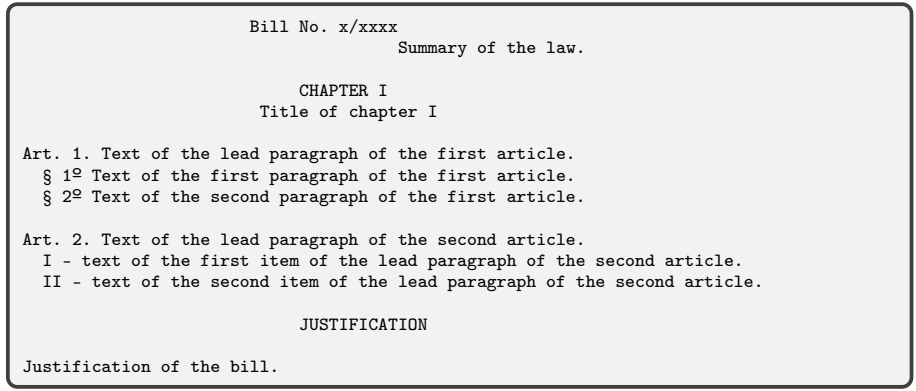

\centering
\begin{tcolorbox}[
    width=0.8\linewidth,
    left=2pt,
    right=2pt,
    top=2pt,
    bottom=2pt,
    halign=flush left
]
\footnotesize
\begin{verbatim}
                       Bill No. x/xxxx
                                      Summary of the law.

                            CHAPTER I
                        Title of chapter I

Art. 1. Text of the lead paragraph of the first article.
  § 1º Text of the first paragraph of the first article.
  § 2º Text of the second paragraph of the first article.

Art. 2. Text of the lead paragraph of the second article.
  I - text of the first item of the lead paragraph of the second article.
  II - text of the second item of the lead paragraph of the second article.

                            JUSTIFICATION

Justification of the bill.
\end{verbatim}
\end{tcolorbox}

\caption{Sample structure of a legislative proposal.}
\label{fig:sample_bill}
\end{figure}

\newcommand{\chunkheader}[3]{%
    \textbf{#1} -- #2\textit{(#3)}
    \par\vspace{2pt}
}

\newcommand{\chunkseparator}{%
    \par\vspace{-10pt}%
    \noindent\rule{\linewidth}{0.4pt}%
    \vspace{2pt}%
}

\begin{figure}[hbp]
\centering
\begin{tcolorbox}[
    width=0.8\linewidth,
    left=2pt,
    right=2pt,
    top=1pt,
    bottom=1pt,
    halign=flush left
]
\footnotesize

\chunkheader{Chunk 1}{Main body of the normative text}{document component level}

\begin{verbatim}
                       Bill No. x/xxxx
                                      Summary of the law.

                            CHAPTER I
                        Title of chapter I

Art. 1. Text of the lead paragraph of the first article.
  § 1º Text of the first paragraph of the first article.
  § 2º Text of the second paragraph of the first article.

Art. 2. Text of the lead paragraph of the second article.
  I - text of the first item of the lead paragraph of the second article.
  II - text of the second item of the lead paragraph of the second article.
\end{verbatim}

\chunkseparator

\chunkheader{Chunk 2}{The justification section}{document component level}

\begin{verbatim}
                            JUSTIFICATION

Justification of the bill.
\end{verbatim}

\chunkseparator

\chunkheader{Chunk 3}{Chapter I}{basic unit hierarchy level}

\begin{verbatim}
                            CHAPTER I
                        Title of chapter I

Art. 1. Text of the lead paragraph of the first article.
  § 1º Text of the first paragraph of the first article.
  § 2º Text of the second paragraph of the first article.

Art. 2. Text of the lead paragraph of the second article.
  I - text of the first item of the lead paragraph of the second article.
  II - text of the second item of the lead paragraph of the second article.
\end{verbatim}

\chunkseparator

\chunkheader{Chunk 4}{Article 1}{basic unit level}

\begin{verbatim}
Art. 1. Text of the lead paragraph of the first article.
  § 1º Text of the first paragraph of the first article.
  § 2º Text of the second paragraph of the first article.
\end{verbatim}

\chunkseparator

\chunkheader{Chunk 5}{Lead paragraph of Article 1}{basic unit component level}

\begin{verbatim}
Art. 1. Text of the lead paragraph of the first article.
\end{verbatim}

\chunkseparator

\chunkheader{Chunk 6}{Paragraph 1 of Article 1}{basic unit component level}

\begin{verbatim}
§ 1º Text of the first paragraph of the first article.
\end{verbatim}

\chunkseparator

\chunkheader{Chunk 7}{Paragraph 2 of Article 1}{basic unit component level}

\begin{verbatim}
§ 2º Text of the second paragraph of the first article.
\end{verbatim}

\chunkseparator

\chunkheader{Chunk 8}{Article 2}{basic unit level}

\begin{verbatim}
Art. 2. Text of the lead paragraph of the second article.
  I - text of the first item of the lead paragraph of the second article.
  II - text of the second item of the lead paragraph of the second article.
\end{verbatim}

\chunkseparator

\chunkheader{Chunk 9}{Item I of Article 2}{enumeration level}

\begin{verbatim}
Art. 2. Text of the lead paragraph of the second article.
  I - text of the first item of the lead paragraph of the second article.
\end{verbatim}

\chunkseparator

\chunkheader{Chunk 10}{Item II of Article 2}{enumeration level}

\begin{verbatim}
Art. 2. Text of the lead paragraph of the second article.
  II - text of the second item of the lead paragraph of the second article.
\end{verbatim}

\end{tcolorbox}

\caption{Chunks generated by segmentation of the sample legal act of Figure \ref{fig:sample_bill}. Each chunk corresponds to a structural unit at a different level of the document hierarchy.}
\label{fig:segmentation_sample_bill}
\end{figure}

When applied to the two test collections, the segmentation procedure does not succeed for all documents. In NormasTCU, 534 documents (3.69\% of the collection) do not follow the required hierarchical structure. The same occurs for 34,380 documents in Ulysses-RFCorpus, which also contains 126 additional documents that, despite following the standard structure, contain structural deviations that prevent the parser from identifying their hierarchical components. Thus, 34,506 Ulysses-RFCorpus documents (32.65\% of the collection) cannot be represented using the defined hierarchy. All documents that cannot be segmented are represented by a single retrieval unit containing their full text, so that every document has a representation in every index.

\subsection{Document representation strategies}
\label{sec:doc_rep_strategies}

Using the multi-layered segmentation described in Section \ref{sec:doc_structural_segmentation}, we evaluate four document representation strategies based on two dimensions: (i) which chunks are indexed and (ii) what context is added to each chunk.

The first strategy, \textit{Hierarchical Chunks}, corresponds to the multi-layered representation proposed by Lima \cite{lima2024unlockinglegalknowledgemultilayered}. It includes all chunks generated by the segmentation (Chunks 1--10 in Figure \ref{fig:segmentation_sample_bill}), with each chunk enriched with the document title, summary, and titles of enclosing structural units.

In Figure~\ref{fig:segmentation_sample_bill}, Chunks 1--3 lack enclosing structural units and are therefore enriched only with the document title and summary, which we refer to as the document \textit{metadata} throughout this paper: the expression \texttt{[Bill No. x/xxxx | Summary of the bill.]} is prepended to these chunks. In contrast, Chunks 4--10 are nested under \textit{Chapter I} and are therefore also enriched with its title: \texttt{[Bill No. x/xxxx | Summary of the bill. | CHAPTER I Title of chapter I]} is prepended to these chunks.

The other three strategies retain only the Basic Unit chunks (articles) and the Document Component chunks that are not part of the normative text, such as the justification of legislative proposals. We refer to these chunks as \textit{core chunks}. In Figure~\ref{fig:segmentation_sample_bill}, the core chunks are Chunks 4 and 8 (Articles 1 and 2) and Chunk 2 (justification). The three strategies differ in the contextual information added to each core chunk.

In the \textit{Core Chunks + Full Context} representation, each core chunk is enriched with the same contextual information used in \textit{Hierarchical Chunks}: the document title, summary, and titles of all enclosing structural units. In our example, Chunk 2 is enriched with the metadata, whereas Chunks 4 and 8 are also enriched with the title of the chapter~I.

The \textit{Core Chunks + Metadata} strategy limits the enrichment to the document metadata. Chunks 2, 4, and 8 are all enriched only with \texttt{[Bill No. x/xxxx | Summary of the Law.]}, and information describing the hierarchical position of the article, such as chapter, section, or subsection titles, is omitted.

Finally, the \textit{Core Chunks + No Context} representation resembles traditional passage segmentation, as each chunk is indexed using only its original text, without any additional context.

Table \ref{tab:representation_strategies} summarizes the strategies.

\begin{table}[!htbp]
\caption{Document representation strategies evaluated in this work.}
\begin{tabular}{lll}
\toprule
\textbf{Strategy} & \textbf{Chunks} & \textbf{Context} \\
\midrule
Hierarchical Chunks & All & Title + Summary + Hierarchy \\
Core Chunks + Full Context & Core & Title + Summary + Hierarchy \\
Core Chunks + Metadata & Core & Title + Summary \\
Core Chunks + No Context & Core & None \\
\bottomrule
\end{tabular}
\label{tab:representation_strategies}
\end{table}

\subsection{Retrieval Pipeline}

For each test collection and representation strategy, we build a separate index using BM25 and three dense retrieval models: OpenAI \texttt{text-embedding-3-small}, \texttt{Qwen3-Embedding-4B}, and \texttt{jua-4B-mixed}.

We selected these models based on previous evaluations on Brazilian legal test collections. BM25 achieved the highest nDCG@10 on both NormasTCU and Ulysses-RFCorpus in the Juá Benchmark \cite{pereira2026juabenchmarkinformation}\footnote{\url{https://huggingface.co/spaces/ufca-llms/jua-leaderboard}}. Among the dense retrievers evaluated in the same benchmark, \texttt{Qwen3-Embedding-4B} achieved the highest MRR on these collections \cite{pereira2026juabenchmarkinformation}. We also include \texttt{jua-4B-mixed}, a domain-adapted embedding model developed for Brazilian legal information retrieval that has achieved state-of-the-art results on some Brazilian legal test collections \cite{pereira2026domainadaptivedenseretrievalbrazilian}. Finally, we include \texttt{text-embedding-3-small} because it has also shown competitive performance on some Brazilian legal retrieval tasks \cite{fernandes2026juristcu}.

Each document $D_i$ is represented by a set of chunks based on the representation strategies described in Section~\ref{sec:doc_rep_strategies}: $D_i=\{c_{i,1},c_{i,2},\ldots,c_{i,n_i}\}$, where $c_{i,j}$ denotes the $j$-th chunk of document $D_i$ and $n_i$ is the number of chunks in the document. During indexing, every chunk is treated as an independent retrieval unit.

Given a query $q$, the retrieval model scores each chunk: $score(q, c_{i,j})$. To map the retrieved chunks to their source document, we define the document score as the highest score among its retrieved chunks: $score(q,D_i)=\max\{ score(q,c_{i_j}) \}$ \cite{parade}. Documents are then ranked according to their scores.

\subsection{Evaluation}

We evaluate each representation strategy on the document-level rankings for NormasTCU and Ulysses-RFCorpus. Both collections provide graded relevance judgments on a three-level scale. We report two metrics: normalized Discounted Cumulative Gain at rank 10 (nDCG@10) and Mean Reciprocal Rank (MRR). nDCG@10 measures the quality of the top-ranked results, considering both the graded relevance of retrieved documents and their positions in the ranking \cite{ndcg}. MRR complements this analysis by measuring how early the first relevant document appears in the ranked list; since MRR requires binary relevance judgments, we consider as relevant any document with a relevance score greater than zero.

\section{Results}
\label{sec:results}

Since each document is represented by one or more chunks, structural segmentation substantially increases the index size (Table~\ref{tab:index_size}). Compared with whole-document indexing, the \textit{Hierarchical Chunks} representation increases the number of indexed units by a factor of ~13 on NormasTCU (from 14,469 to 192,955) and ~9 on Ulysses-RFCorpus (from 105,669 to 971,044). Even the \textit{Core Chunks} representation indexes considerably more units than the whole-document baseline, by a factor of ~5 on NormasTCU and ~4 on Ulysses-RFCorpus. Therefore, any performance gains stem from better text representation rather than from a smaller index.

\begin{table}[!htbp]
\centering
\caption{Number of indexed retrieval units for each document representation strategy.}
\label{tab:index_size}
\begin{tabular}{lcc}
\hline
\textbf{Representation} & \textbf{NormasTCU} & \textbf{Ulysses-RFCorpus} \\
\hline
Whole Document & 14,469 & 105,669 \\
Hierarchical Chunks & 192,955 & 971,044 \\
Core Chunks & 78,157 & 417,690 \\
\hline
\end{tabular}
\end{table}

Table~\ref{tab:resultados} summarizes retrieval performance across strategies. The first block reports the whole-document results from the Juá benchmark \cite{pereira2026juabenchmarkinformation}, which serve as the baseline for each retriever. Two aspects of these baselines are worth noting. First, BM25 is the strongest baseline in terms of nDCG@10 on both collections, confirming the competitiveness of lexical retrieval on Brazilian legal collections that motivated this work. Second, the baselines are considerably stronger on Ulysses-RFCorpus than on NormasTCU for all retrievers, leaving less room for improvement through segmentation (e.g., for BM25, nDCG@10 is 0.364 on NormasTCU and 0.531 on Ulysses-RFCorpus).

\begin{table}[htbp]
\centering
\caption{Retrieval results on the NormasTCU and Ulysses-RFCorpus datasets. Baseline extracted from \cite{pereira2026juabenchmarkinformation, pereira2026domainadaptivedenseretrievalbrazilian}}
\renewcommand{\arraystretch}{1.2}
\setlength{\tabcolsep}{10pt}
\begin{tabular}{lcccc}
\hline
\multirow{2}{*}{\textbf{Retriever}} & \multicolumn{2}{c}{\textbf{NormasTCU}} & \multicolumn{2}{c}{\textbf{Ulysses-RFCorpus}} \\
\cmidrule(lr){2-3}\cmidrule(lr){4-5}
 & nDCG@10 & MRR & nDCG@10 & MRR \\
\hline

\multicolumn{5}{c}{\textit{Baseline: whole-document indexing (Juá Benchmark)}} \\
\hline
BM25 & 0.364 & 0.444 & 0.531 & 0.696 \\
text-embedding-3-small & 0.263 & 0.427 & 0.411 & 0.583 \\
jua-4B-mixed & 0.305 & 0.474 & 0.441 & 0.624 \\
Qwen3-Embedding-4B & 0.307 & 0.508 & 0.450 & \textbf{0.719} \\

\hline
\multicolumn{5}{c}{\textit{Hierarchical Chunks}} \\
\hline
BM25 & 0.342 & 0.509 & 0.532 & 0.700 \\
text-embedding-3-small & 0.358 & 0.555 & 0.456 & 0.621 \\
jua-4B-mixed & 0.355 & 0.540 & 0.482 & 0.649 \\
Qwen3-Embedding-4B & 0.333 & 0.517& 0.511 & 0.668 \\

\hline
\multicolumn{5}{c}{\textit{Core Chunks + Full Context}} \\
\hline
BM25 & 0.332 & 0.524 & \textbf{0.535} & 0.702 \\
text-embedding-3-small & 0.342 & 0.525 & 0.451 & 0.610 \\
jua-4B-mixed & 0.383 & \textbf{0.566} & 0.488 & 0.655 \\
Qwen3-Embedding-4B & 0.351 & 0.537 & 0.515 & 0.674 \\

\hline
\multicolumn{5}{c}{\textit{Core Chunks + Metadata}} \\
\hline
BM25 & 0.341 & 0.507 & 0.529 & 0.696 \\
text-embedding-3-small & 0.354 & 0.520 & 0.448 & 0.607 \\
jua-4B-mixed & \textbf{0.392} & 0.558 & 0.484 & 0.654 \\
Qwen3-Embedding-4B & 0.340 & 0.522 & 0.508 & 0.672 \\

\hline
\multicolumn{5}{c}{\textit{Core Chunks + No Context}} \\
\hline
BM25 & 0.255 & 0.430 & 0.490 & 0.676 \\
text-embedding-3-small & 0.301 & 0.502 & 0.418 & 0.580 \\
jua-4B-mixed & 0.297 & 0.435 & 0.475 & 0.632 \\
Qwen3-Embedding-4B & 0.269 & 0.406 & 0.473 & 0.632 \\
\hline

\end{tabular}
\label{tab:resultados}
\end{table}

\textit{Hierarchical Chunks} boosts nDCG@10 for the three dense retrievers on both collections. On NormasTCU, the largest relative gains occur for retrievers with the weakest baselines: 36\% for \texttt{text-embedding-3-small} (from 0.263 to 0.358), 16\% for \texttt{jua-4B-mixed} (0.305 to 0.355), and 8\% for \texttt{Qwen3-Embedding-4B} (0.307 to 0.333). On Ulysses-RFCorpus, the relative gains are more uniform, ranging from 9\% to 14\% across the three dense retrievers. For BM25, nDCG@10 decreases on NormasTCU (from 0.364 to 0.342) and remains almost unchanged on Ulysses-RFCorpus (from 0.531 to 0.532), i.e., lexical retrieval does not benefit from segmentation in the same way as dense retrieval.

The simplified core-based representations with context achieve results comparable to or better than \textit{Hierarchical Chunks} while indexing fewer units. With \textit{Core Chunks + Full Context}, nDCG@10 is close to that of \textit{Hierarchical Chunks} on Ulysses-RFCorpus (absolute differences of at most 0.006). On NormasTCU, it outperforms \textit{Hierarchical Chunks} with two of the three dense retrievers, yielding gains of +0.028 for \texttt{jua-4B-mixed} and +0.018 for \texttt{Qwen3-Embedding-4B}, while underperforming by $-$0.016 with \texttt{text-embedding-3-small}. \textit{Core Chunks + Metadata} shows a similar pattern, with differences of at most 0.008 on Ulysses-RFCorpus. On NormasTCU, this strategy achieves the highest nDCG@10 among all evaluated dense retrievers, reaching 0.392 with \texttt{jua-4B-mixed}.

Removing the context (\textit{Core Chunks + No Context}) lowers nDCG@10 for all retrievers on both collections compared with \textit{Core Chunks + Full Context} and \textit{Core Chunks + Metadata}. The comparison with the whole-document baselines is less consistent: without context, segmentation still outperforms the baselines for all dense retrievers on Ulysses-RFCorpus, but only for \texttt{text-embedding-3-small} on NormasTCU. It also substantially reduces BM25 performance on both collections, from 0.364 to 0.255 on NormasTCU and from 0.531 to 0.490 on Ulysses-RFCorpus.

The MRR followed a similar trend. \textit{Hierarchical Chunks} increases MRR compared with whole-document indexing in almost all cases, except for \texttt{Qwen3-Embedding-4B} on Ulysses-RFCorpus. The \textit{Core Chunks + Full Context} and \textit{Core Chunks + Metadata} achieve similar results to \textit{Hierarchical Chunks} on Ulysses-RFCorpus and further improve MRR for \texttt{jua-4B-mixed} and \texttt{Qwen3-Embedding-4B}. As with nDCG@10, \textit{Core Chunks + No Context} produces lower MRR than \textit{Core Chunks + Full Context} and \textit{Core Chunks + Metadata} in all cases.

Two cases stand out in the MRR results. The first occurs on NormasTCU with BM25: although segmentation reduces nDCG@10, MRR increases from 0.444 with whole-document indexing to 0.509 with \textit{Hierarchical Chunks} and 0.524 with \textit{Core Chunks + Full Context}. Thus, the first relevant document tends to appear earlier in the ranked list even though overall top-10 ranking quality does not improve. The second case occurs on Ulysses-RFCorpus with \texttt{Qwen3-Embedding-4B}: MRR decreases from 0.719 (baseline) to 0.668 (\textit{Hierarchical Chunks}). For this metric and retriever, none of the evaluated representation strategies outperforms the whole-document result on Ulysses-RFCorpus.

\begin{figure}[hbp]
    \centering
    \includegraphics[width=.92\linewidth]{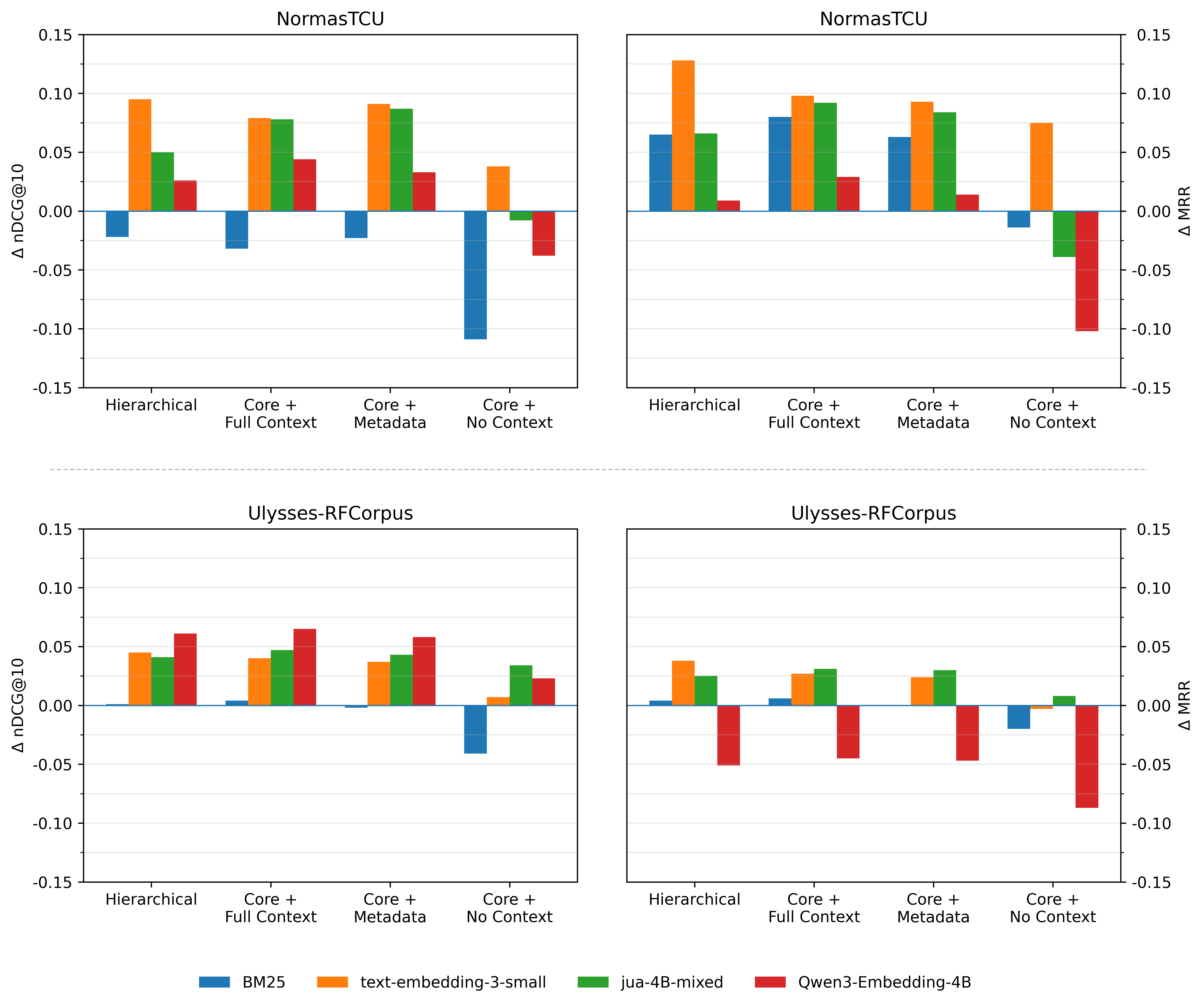}
    \caption{Change ($\Delta$) in retrieval effectiveness relative to the corresponding whole-document baseline on NormasTCU (top row) and Ulysses-RFCorpus (bottom row), for nDCG@10 (left column) and MRR (right column). For each retriever, $\Delta$ is computed relative to its whole-document result.}
    \label{fig:delta_results}
\end{figure}

Figure~\ref{fig:delta_results} summarizes these results by showing the change in nDCG@10 and MRR relative to the corresponding whole-document baseline. For each strategy, retriever, and collection, we compute the difference relative to its whole-document result. Thus, each value represents the gain or loss from changing only the document representation while keeping the retrieval model fixed. The figure shows several patterns:

\begin{itemize}
    \item Dense retrievers consistently improve nDCG@10 with all three context-enriched representations. The same holds for MRR, except for \texttt{Qwen3-Embedding-4B} on Ulysses-RFCorpus, where the whole-document baseline remains the best result;
    \item The \textit{Core Chunks + Full Context} and \textit{Core Chunks + Metadata} achieve results close to \textit{Hierarchical Chunks} on Ulysses-RFCorpus and tend to outperform it on NormasTCU;
    \item \textit{Core Chunks + No Context} performs worse than \textit{Core Chunks + Full Context} and \textit{Core Chunks + Metadata};
    \item The results for BM25 are less consistent: nDCG@10 and MRR move in opposite directions on NormasTCU; both metrics remain almost stable on Ulysses-RFCorpus, except for a decline with the \textit{Core Chunks + No Context} representation.
\end{itemize}

\section{Discussion of the Results and Practical Implications}

\subsection{Discussion of the Results}

Overall, the gains from segmentation are larger on NormasTCU than on Ulysses-RFCorpus. The characteristics of Ulysses-RFCorpus may help explain this difference. Its whole-document baselines are already strong, leaving less room for improvement through segmentation. This may be related to both the documents and the queries. On the document side, each legislative proposal includes a justification in which the deputy describes the problem to be solved and the proposed solution. This section semantically enriches the document and mitigates the vocabulary mismatch problem of lexical retrieval. On the query side, deputies use natural-language descriptions of the intended legislative proposal. In contrast, NormasTCU has no justification sections and usually has short keyword queries whose terms may not appear in the normative text.

The gains on Ulysses-RFCorpus are also limited by the documents that do not follow the expected hierarchical structure. Approximately one-third of the collection cannot be segmented and is therefore represented as a single indexed unit. For these documents, the segmented indexes behave like whole-document indexing, reducing the potential gains from segmentation on this collection.

The results allow us to answer the proposed research questions:

\textbf{RQ1.} Does structural segmentation improve Brazilian legal document retrieval compared with whole-document indexing?

\textit{Answer:} For dense retrievers, the three representations that include contextual information (\textit{Hierarchical Chunks}, \textit{Core Chunks + Full Context}, and \textit{Core Chunks + Metadata}) outperform whole-document indexing, whereas \textit{Core Chunks + No Context} does not. Structural segmentation, however, does not consistently benefit BM25.

\textbf{RQ2.} Which representation components contribute most to retrieval effectiveness?

\textit{Answer:} For dense retrievers, representations based on core chunks with contextual enrichment use substantially fewer indexed units than \textit{Hierarchical Chunks} while achieving comparable or better retrieval results in most cases. We found no clear evidence that full hierarchical context outperforms core chunks enriched with document metadata (title and summary). We therefore recommend the simpler \textit{Core Chunks + Metadata} representation.  

\subsection{Practical Implications}

The results suggest that legal retrieval systems can use the structure of legal acts without indexing all levels of their hierarchy. The \textit{Core Chunks + Metadata} representation achieved results comparable to or better than \textit{Hierarchical Chunks} while indexing substantially fewer units. They also show that contextual information is important, as removing document metadata consistently reduced retrieval effectiveness. Thus, practical implementations can focus on core structural units, such as articles, enriched with concise document-level metadata.

\section{Limitations of the study}\label{sec:limitations}

This study has four main limitations. First, we evaluated only two Brazilian legal test collections, NormasTCU and Ulysses-RFCorpus. They are the only available datasets that match the characteristics required by this study and cover different types of legal documents and query formulations, but they do not represent the full diversity of Brazilian legal documents. Moreover, about one-third of the Ulysses-RFCorpus documents do not follow the expected structure and therefore cannot be represented using the proposed hierarchical segmentation.

Second, we evaluated BM25 and three dense embedding models using max-score aggregation only. Our findings may not generalize to other retrieval models or chunk aggregation methods.

Third, we tested four representation strategies based on  Lima \cite{lima2024unlockinglegalknowledgemultilayered}. Although our experiment show the effects of core chunks and contextual enrichment, we dit not test every possible combination of structural components and contextual information.

Finally, our evaluation focuses on retrieval effectiveness measured by nDCG@10 and MRR. These metrics capture different aspects of ranking quality, but do not measure other practical dimensions of a retrieval system, such as storage or computational costs.

\section{Conclusions}\label{sec:conclusions}

We investigated whether structural segmentation improves the retrieval of Brazilian legal documents. We tested four document representations based on their multi-layered structure: one hierarchical representation and three simplified representations based on core chunks with different levels of context.

The results show that context-enriched representations improve dense retrieval over whole-document indexing. On the evaluated test collections, \textit{Hierarchical Chunks} increased nDCG@10 for all three dense retrievers relative to whole-document indexing, despite substantially increasing the number of indexed units. The same pattern was generally observed for MRR. For BM25, however, structural segmentation did not improve nDCG@10, although it improved MRR in some configurations.

Ablation studies show that the complete hierarchical representation is not necessary to achieve these gains. The \textit{Core Chunks + Full Context} and \textit{Core Chunks + Metadata} representations achieved results close to -- and in some cases better than -- \textit{Hierarchical Chunks}, while indexing substantially fewer indexed units. In contrast, removing context consistently reduced retrieval effectiveness.

Overall, structural segmentation effectively boosts Brazilian legal information retrieval.  However, the complete structure of a legal act is not necessary to achieve these gains, and a simplified representation based on core chunks enriched with document metadata offers a favorable trade-off between representation complexity and retrieval effectiveness. Future work should evaluate additional structural representations, retrieval models, and aggregation methods, including segmentation strategies for legal documents that do not follow the structure considered in this article.

%\section*{Funding}
%The Article Processing Charge (APC) for the publication of this research was funded by the Coordenação de Aperfeiçoamento de Pessoal de Nível Superior - Brasil (CAPES) (ROR identifier: 00x0ma614). For open access purposes, the authors have applied the Creative Commons CC BY license to any accepted version of the article.

\section*{Declaration of generative AI and AI-assisted technologies in the manuscript preparation process}
The authors used AI-assisted tools for language revision during manuscript preparation. All technical content, interpretation, and final editorial decisions were reviewed by the authors, who take full responsibility for the article.

\bibliography{bibliography}

@article{fernandes2026juristcu,
  author    = {Leandro Carísio Fernandes and 
               Leandro dos Santos Ribeiro and 
               Marcos Vinícius Borela de Castro and 
               Leonardo Augusto da Silva Pacheco and 
               Edans Flávius de Oliveira Sandes},
  title     = {{JurisTCU}: a {B}razilian {P}ortuguese information retrieval dataset with query relevance judgments},
  journal   = {Language Resources and Evaluation},
  year      = {2026},
  volume    = {60},
  number    = {1},
  doi       = {10.1007/s10579-025-09881-w},
  issn      = {1574-0218}
}

@article{normastcu,
    title={{NormasTCU} — A {B}razilian {P}ortuguese {IR} Dataset and an Evaluation of {LLM}-as-a-Judge for Relevance Assessment}, 
    author={Leandro Carísio Fernandes and Marcus Vinícius Borela de Castro and Leandro dos Santos Ribeiro and Leonardo Augusto da Silva Pacheco and Edans Flávius de Oliveira Sandes},
    journal = {Information Processing \& Management},
    volume = {64},
    number = {2, Part A},
    year = {2027},
    issn = {0306-4573},
    doi={10.1016/j.ipm.2026.105151}
}

@misc{pereira2026juabenchmarkinformation,
      title={{JU\'A} -- A Benchmark for Information Retrieval in {B}razilian Legal Text Collections}, 
      author={Jayr Pereira and Leandro Carísio Fernandes and Erick de Brito and Roberto Lotufo and Luiz Bonifacio},
      year={2026},
      eprint={2604.06098},
      archivePrefix={arXiv},
      primaryClass={cs.IR},
      doi={10.48550/arXiv.2604.06098}
}

@InProceedings{callan_passage,
  author="Callan, James P.",
  editor="Croft, Bruce W.
  and van Rijsbergen, C. J.",
  title={Passage-Level Evidence in Document Retrieval},
  booktitle="SIGIR '94",
  year="1994",
  publisher="Springer London",
  address="London",
  pages="302--310",
  isbn="978-1-4471-2099-5"
}

@inproceedings{passage_retrieval_lm,
  author = {Liu, Xiaoyong and Croft, W. Bruce},
  title = {Passage retrieval based on language models},
  year = {2002},
  isbn = {1581134924},
  publisher = {Association for Computing Machinery},
  address = {New York, NY, USA},
  doi = {10.1145/584792.584854},
  booktitle = {Proceedings of the Eleventh International Conference on Information and Knowledge Management},
  pages = {375–382},
  numpages = {8},
  location = {McLean, Virginia, USA},
  series = {CIKM '02}
}

@article{ulyssesrfcorpus,
  author    = {Douglas Vitório and Ellen Souza and Lucas Martins and Nádia F. F. da Silva and André Carlos Ponce de Leon de Carvalho and Adriano L. I. Oliveira and Francisco Edmundo de Andrade},
  title     = {Building a relevance feedback corpus for legal information retrieval in the real-case scenario of the {B}razilian {C}hamber of {D}eputies},
  journal   = {Language Resources and Evaluation},
  year      = {2024},
  volume    = {},
  number    = {},
  pages     = {},
  month     = aug,
  doi       = {10.1007/s10579-024-09767-3},
  issn      = {1574-0218}
}

@inproceedings{brtax,
  author = {J\'{u}nior, Juvenal Domingos and Faria, Augusto and de Oliveira, E. Seiti and de Brito, Erick and Teotonio, Matheus and Assump\c{c}\~{a}o, Andre and Carmo, Diedre and Lotufo, Roberto and Pereira, Jayr},
  title = {{BR-TaxQA-R}: A Dataset for Question Answering with References for {B}razilian Personal Income Tax Law, Including Case Law},
  year = {2025},
  isbn = {978-3-032-15986-1},
  publisher = {Springer-Verlag},
  address = {Berlin, Heidelberg},
  doi = {10.1007/978-3-032-15987-8_14},
  booktitle = {Intelligent Systems: 35th Brazilian Conference, BRACIS 2025, Fortaleza-CE, Brazil, September 29 – October 2, 2025, Proceedings, Part I},
  pages = {208-222},
  numpages = {15},
  location = {Fortaleza-CE, Brazil}
}

@article{Moens2001,
  author    = {Marie-Francine Moens},
  title     = {Innovative Techniques for Legal Text Retrieval},
  journal   = {Artificial Intelligence and Law},
  year      = {2001},
  volume    = {9},
  number    = {1},
  pages     = {29--57},
  doi       = {10.1023/A:1011297104922},
  issn      = {1572-8382},
  month     = mar
}

@misc{pipeline_cd,
  title   = {An information retrieval pipeline for legislative documents from the {B}razilian {C}hamber of {D}eputies},
  author = {Souza, Ellen Polliana Ramos and Vitório, Douglas and Moriyama, Gyovana and Santos, Luiz and Martins, Lucas and Souza, Mariana and Fonseca, Márcio and Silva, Nádia Felix Felipe da and Carvalho, André Carlos Ponce de Leon Ferreira de and Albuquerque, Hidelberg Oliveira and Oliveira, Adriano Lorena Inacio de},
  year = {2021},
  doi = {10.3233/FAIA210326},
  publisher = {IOS Press},
  booktitle   = {{Annual Conference on Legal Knowledge and Information Systems - JURIX}}
}

@misc{lima2024unlockinglegalknowledgemultilayered,
  title={Unlocking Legal Knowledge with Multi-Layered Embedding-Based Retrieval},
  author={João Alberto de Oliveira Lima},
  year={2024},
  eprint={2411.07739},
  archivePrefix={arXiv},
  primaryClass={cs.AI},
  doi={10.48550/arXiv.2411.07739}
}

@misc{demartim2026probabilisticsimilaritystructuraltemporal,
      title={Beyond Probabilistic Similarity: Structural, Temporal, and Causal Limitations of Retrieval-Augmented Generation in the Legal Domain}, 
      author={Hudson de Martim},
      year={2026},
      eprint={2606.09724},
      archivePrefix={arXiv},
      primaryClass={cs.AI},
      doi={10.48550/arXiv.2606.09724}, 
}

@article{parade,
    author = {Li, Canjia and Yates, Andrew and MacAvaney, Sean and He, Ben and Sun, Yingfei},
    title = {{PARADE}: Passage Representation Aggregation for Document Reranking},
    year = {2023},
    issue_date = {March 2024},
    publisher = {Association for Computing Machinery},
    address = {New York, NY, USA},
    volume = {42},
    number = {2},
    issn = {1046-8188},
    doi = {10.1145/3600088},
    journal = {ACM Trans. Inf. Syst.},
    month = sep,
    articleno = {36},
    numpages = {26}
}

@misc{zhao2022densetextretrievalbased,
      title={Dense Text Retrieval based on Pretrained Language Models: A Survey},
      author={Wayne Xin Zhao and Jing Liu and Ruiyang Ren and Ji-Rong Wen},
      year={2022},
      eprint={2211.14876},
      archivePrefix={arXiv},
      primaryClass={cs.IR},
      doi={10.48550/arXiv.2211.14876}
}

@InProceedings{Ai_2018,
    author="Ai, Qingyao
    and O'Connor, Brendan
    and Croft, W. Bruce",
    editor="Pasi, Gabriella
    and Piwowarski, Benjamin
    and Azzopardi, Leif
    and Hanbury, Allan",
    title={A Neural Passage Model for Ad-hoc Document Retrieval},
    booktitle="Advances in Information Retrieval",
    year="2018",
    publisher="Springer International Publishing",
    address="Cham",
    pages="537--543",
    isbn="978-3-319-76941-7"
}

@misc{Brasil1998LC95,
  author     = {{Brasil}},
  title      = {{Lei Complementar nº 95, de 26 de fevereiro de 1998}},
  year       = {1998}
}

@misc{Brasil2024Dec12002,
  author     = {{Brasil}},
  title      = {{Decreto nº 12.002, de 22 de abril de 2024}},
  year       = {2024}
}

@misc{Brasil2003ResTCU164,
  author     = {{Brasil}},
  title      = {{Resolução-TCU nº 164, de 8 de outubro de 2003}},
  year       = {2003}
}

@misc{pereira2026domainadaptivedenseretrievalbrazilian,
      title={Domain-Adaptive Dense Retrieval for {B}razilian Legal Search}, 
      author={Jayr Pereira and Roberto Lotufo and Luiz Bonifacio},
      year={2026},
      eprint={2605.04005},
      archivePrefix={arXiv},
      primaryClass={cs.IR},
      doi={10.48550/arXiv.2605.04005}
}

@article{ndcg,
    author = {J\"{a}rvelin, Kalervo and Kek\"{a}l\"{a}inen, Jaana},
    title = {Cumulated gain-based evaluation of {IR} techniques},
    year = {2002},
    issue_date = {October 2002},
    publisher = {Association for Computing Machinery},
    address = {New York, NY, USA},
    volume = {20},
    number = {4},
    issn = {1046-8188},
    doi = {10.1145/582415.582418},
    journal = {ACM Trans. Inf. Syst.},
    month = oct,
    pages = {422-446},
    numpages = {25}
}

@article{rosso2011passage,
    title = {Passage retrieval in legal texts},
    journal = {The Journal of Logic and Algebraic Programming},
    volume = {80},
    number = {3},
    pages = {139-153},
    year = {2011},
    issn = {1567-8326},
    doi = {10.1016/j.jlap.2011.02.001},
    author = {Paolo Rosso and Santiago Correa and Davide Buscaldi},
}

@inproceedings{aumiller2021structural,
   series={ICAIL '21},
   title={Structural text segmentation of legal documents},
   DOI={10.1145/3462757.3466085},
   booktitle={Proceedings of the Eighteenth International Conference on Artificial Intelligence and Law},
   publisher={ACM},
   author={Aumiller, Dennis and Almasian, Satya and Lackner, Sebastian and Gertz, Michael},
   year={2021},
   month=June,
   pages={2–11},
   collection={ICAIL ’21}
}

@inproceedings{malik2022semantic,
    title = {Semantic Segmentation of Legal Documents via Rhetorical Roles},
    author = "Malik, Vijit  and
      Sanjay, Rishabh  and
      Guha, Shouvik Kumar  and
      Hazarika, Angshuman  and
      Nigam, Shubham Kumar  and
      Bhattacharya, Arnab  and
      Modi, Ashutosh",
    booktitle = "Proceedings of the Natural Legal Language Processing Workshop 2022",
    month = dec,
    year = "2022",
    address = "Abu Dhabi, United Arab Emirates (Hybrid)",
    publisher = "Association for Computational Linguistics",
    doi = "10.18653/v1/2022.nllp-1.13",
    pages = "153--171",
}

@inproceedings{li2023sailer,
    author = {Li, Haitao and Ai, Qingyao and Chen, Jia and Dong, Qian and Wu, Yueyue and Liu, Yiqun and Chen, Chong and Tian, Qi},
    title = {{SAILER}: Structure-aware Pre-trained Language Model for Legal Case Retrieval},
    year = {2023},
    isbn = {9781450394086},
    publisher = {Association for Computing Machinery},
    address = {New York, NY, USA},
    doi = {10.1145/3539618.3591761},
    booktitle = {Proceedings of the 46th International ACM SIGIR Conference on Research and Development in Information Retrieval},
    pages = {1035–1044},
    numpages = {10},
    location = {Taipei, Taiwan},
    series = {SIGIR '23}
}

@article{YANG2026104816,
    title = {LLM-guided multi-view representation learning for legal case retrieval via structured event chains and role-aware aggregation},
    journal = {Information Processing \& Management},
    volume = {63},
    number = {7, Part A},
    pages = {104816},
    year = {2026},
    issn = {0306-4573},
    doi = {10.1016/j.ipm.2026.104816},
    author = {Caiwei Yang and Wei Gao and Shuai Yu and Yanping Chen and Yongbin Qin and Ruizhang Huang and Chuan Lin}
}

@article{BHATTACHARYA2022103069,
    title = {Legal case document similarity: You need both network and text},
    journal = {Information Processing \& Management},
    volume = {59},
    number = {6},
    pages = {103069},
    year = {2022},
    issn = {0306-4573},
    doi = {10.1016/j.ipm.2022.103069},
    author = {Paheli Bhattacharya and Kripabandhu Ghosh and Arindam Pal and Saptarshi Ghosh}
}

@article{NGUYEN2025103949,
    title = {Retrieve–Revise–Refine: A novel framework for retrieval of concise entailing legal article set},
    journal = {Information Processing \& Management},
    volume = {62},
    number = {1},
    pages = {103949},
    year = {2025},
    issn = {0306-4573},
    doi = {10.1016/j.ipm.2024.103949},
    author = {Chau Nguyen and Phuong Nguyen and Le-Minh Nguyen}
}
%\printbibliography

\end{document}